\documentclass[12pt,preprint]{aastex}
\def\gtorder{\mathrel{\raise.3ex\hbox{$>$}\mkern-14mu
                \lower0.6ex\hbox{$\sim$}}}
\def\ltorder{\mathrel{\raise.3ex\hbox{$<$}\mkern-14mu
                \lower0.6ex\hbox{$\sim$}}}

\shorttitle{CN and HCN in PDRs}
\shortauthors{Boger \& Sternberg}

\begin{document}
\title{CN and HCN in Dense Interstellar Clouds}
\vspace{1cm}
\author{Gai I. Boger and Amiel Sternberg}
\vspace{0.5cm}
\affil{School of Physics and Astronomy and the Wise Observatory,
        The Beverly and Raymond Sackler Faculty of Exact Sciences,
        Tel Aviv University, Tel Aviv 69978, Israel}
\email{amiel@wise.tau.ac.il}

\begin{abstract}
We present a theoretical investigation of CN and HCN molecule formation 
in dense interstellar clouds.  
We study the gas-phase 
CN and HCN production efficiencies
from the outer  photon-dominated regions (PDRs) into the opaque cosmic-ray
dominated cores. We calculate the equilibrium densities
of CN and HCN, and of the associated species C$^+$, C, and CO,
as functions of the far-ultraviolet (FUV) optical depth.
We consider isothermal gas at 50 K, with 
hydrogen particle densities from $10^2$ to $10^6$ cm$^{-3}$.
We study clouds that are exposed to 
FUV fields with intensities (at 1000 \AA) from
$5\times 10^{-19}$ to $5\times 10^{-14}$ 
erg~s$^{-1}$~cm$^{-2}$~Hz~sr$^{-1}$, or 
$20$ to $2\times 10^5$ times the mean interstellar FUV intensity.
We assume cosmic-ray H$_2$ ionization rates ranging from 
$5\times 10^{-17}$~s$^{-1}$, to an enhanced value of
$5\times 10^{-16}$~s$^{-1}$.
We also examine the sensitivity of the density profiles to
the gas-phase sulfur abundance.

\end{abstract}



\section{Introduction}
\label{introduction}
 
Millimeter-wave line emissions of
CN and HCN molecules are widely used probes of 
dense molecular gas and of photon-dominated regions (PDRs)
in the Galactic interstellar medium (ISM).
Observed sources include molecular interfaces 
in star-forming regions (Greaves \& Church 1996; Simon et al.~1997; 
Young Owl et al.~2000; Savage et al.~2002; Schneider et al.~2003;  
Johnstone et al.~2003), reflection nebulae (Fuente et al.~1993, 1995, 2003; 
Jansen~et al. 1995), planetary nebulae (Bachiller et al.~1997a), 
and circumstellar envelopes and disks
(Wooten et al.~1982; Truong-Bach et al.~1987; Bachiller et al.~1997b; 
Lindqvist et al.~2000; van Zadelhoff et al.~2003; Thi et al.~2004).
Observations of
CN and HCN have recently been used to study PDRs in the 
starburst galaxy M82 and in other external systems 
(Fuente et al.~2005; Meier \& Turner 2005). 

In nearby objects, such as the Orion Bar 
and the reflection nebulae NGC 2023 and 7023, the
molecular emissions have been mapped across the PDRs.
The CN/HCN intensity ratios are largest near the
stellar sources of the illuminating far-ultraviolet (FUV) radiation fields, 
and the intensity and density ratios decrease with
increasing optical depth and distance from the stars.
This behavior is broadly consistent with theoretical expectations 
(Sternberg \& Dalgarno 1995 [SD95]; Jansen et al.~1995) and
is evidence of selective photodissociation of HCN versus CN 
(van Zadelhoff et al.~2003; Thi et al.~2004). 

Many discussions of gas-phase nitrogen chemistry in molecular clouds
have been presented in the literature 
(e.g., Herbst \& Klemperer 1973; Prasad \& Huntress 1980;
Viala 1986; Pineau des Forets et al.~1990; Herbst et al.~1994; SD95;
Lee et al.~1996; Turner et al.~1997). 
Comprehensive reviews of PDR observations and theory, 
and related subjects have been presented by Hollenbach \& Tielens (1999), 
Sternberg (2004), and van Dishoeck (2005).

In this paper we focus on the gas-phase production of CN and HCN,
and present results 
for a wide range of conditions.
We analyze how the CN and HCN formation
and destruction sequences vary with optical depth, first through the
PDRs and then into the opaque cosmic-ray dominated cores.
We discuss how the (equilibrium) density profiles depend on
the cloud hydrogen gas densities and the incident FUV field intensities, 
and we identify the qualitative
changes in the density profiles that may be expected 
in moving from low- to high-density systems.
The production efficiencies of the carbon bearing
CN and HCN molecules depend
on the availability of free C$^+$ ions and C atoms in the
gas, and we present and discuss the associated 
C$^+$, C, and CO density profiles in the parameter space we consider. 
In this paper we do not consider HNC or the 
isomeric abundance ratio HCN/HNC
(e.g., Watson 1974; Schilke et. al.~1992; Herbst et al.~2000)
which, together with CN/HCN, may be expected to vary with optical
depth, density, and FUV field strength.

In \S 2 we describe the basic ingredients of our models.
In \S 3 we discuss the gas-phase CN and HCN reaction sequences
that operate in molecular clouds.
In \S 4 we present detailed results for
a ``reference model'' to illustrate the depth-dependent
formation pathways and density profiles. 
In \S 5 we present our parameter study and discuss results
for a range of gas densities, FUV field strengths,
cosmic-ray ionization rates, and gas-phase elemental abundances.
We summarize in \S6.


\section{Model Ingredients}

We performed our model computations using an updated version of the SD95 
code.  The models consist of static, plane-parallel, semi-infinite slabs, 
exposed on one side to isotropic FUV (6-13.6 eV) radiation fields. 
The steady-state abundances of the atomic and molecular species are
computed as functions of the visual extinction, $A_V$, from the cloud surface.
The models account for scattering and absorption of the FUV photons
by dust grains, and the resulting depth-dependent attenuation of
the atomic and molecular photodissociation and photoionization rates. 
The effects of H$_2$ and CO absorption-line shielding are also included.
The molecular chemistry is driven by the combined action of
FUV photoionization and photodissociation, and cosmic-ray impact
ionization. The resulting sequences of (two-body) gas-phase
ion-molecule and neutral-neutral reactions are mediated by
dissociative recombination and photo-destruction.

We assume that the spectral shapes of the
incident FUV fields are identical to the Draine (1978) 
representation of the interstellar field in the solar neighborhood (see also
Parravano, Hollenbach \& McKee 2003). The field intensity is
parameterized by a scaling factor $\chi$, where
for $\chi=1$ the FUV intensity at 1000 \AA \ is 
$5.4\times 10^{-20}$ erg s$^{-1}$ cm$^{-2}$ Hz$^{-1}$ sr$^{-1}$. 
We construct models for $\chi$ ranging from $20$ to 
$2\times 10^5$, appropriate for PDRs in the vicinity of
young OB stars and clusters (Sternberg et al.~2003).

In our computations we include the same set of 70 atomic and molecular 
carbon, nitrogen, oxygen, sulfur, and silicon bearing
species 
\footnote{
We assume that the abundances of all other elements are
negligible in the gas phase,
including heavy elements such as Mg and Fe
which, when abundant, may become important positive-charge carriers in
dense molecular clouds
(Oppenheimer \& Dalgarno 1974; de Boisanger et al.~1996).
}
considered by SD95.
Complex hydrocarbons and PAHs are excluded.
We solve the depth dependent equations of chemical equilibrium
\begin{equation}
\sum_{jl}k_{ijl}(T)n_jn_l + \sum_j [\chi\Gamma_{ij}+\xi_{ij}]n_j
= n_i\biggl\lbrace \sum_{jl}k_{jil}n_l + 
\sum_j[\chi\Gamma_{ji}+\xi_{ji}] \biggr\rbrace 
\end{equation}
as functions of $A_V$.
In these equations,
$n_i$ are the densities (cm$^{-3}$) of species $i$,
and $k_{ijl}(T)$ are the (temperature-dependent) rate coefficients 
(cm$^3$ s$^{-1}$) for chemical
reactions between species $j$ and $l$ that lead to the
production of $i$. The parameters $\chi\Gamma_{ij}$ 
and $\xi_{ij}$ are the FUV photon and cosmic-ray destruction
rates (s$^{-1}$) of species $j$ with products $i$.
We solve Equations (1) via Newton-Raphson iteration, and
use Bulirsch-Stoer integration to compute the column densities.

In our computations we include all reactions listed in the
{\small UMIST99} database (Le Teuff et al.~2000) for which the
reactants and products are species in our set.
We adopt the recommended {\small UMIST99} rate-coefficients $k_{ijl}$,
with several alterations and updates 
\footnote{Our input list of rate coefficients is available at
ftp://wise3.tau.ac.il/pub/amiel/pdr. In our compilation
we use recent determinations of
the rate coefficients and branching ratios for 
dissociative recombinations (Vejby-Christensen et al.~1997; 
Larson et al.~1998; Vikor et al.~1999; Jensen et al.~2000;
McCall et al.~2003; Geppert et al.~2004).
For some neutral-neutral reactions we use 
rate-coefficients listed in {\small UMIST95},
or as given by Herbst et al.~(2000) and Smith et al.~(2004).  
We do not assume the small ($\sim 50$ K)
activation barriers postulated by Pineau des Forets et al.~(1990)
for the neutral-neutral reactions (see \S 3) involved in the
nitrogen chemistry.}.

The depth-dependent FUV photodissociation
and photoionization rates are crucial quantities. They depend on
the intensity and spectral shape of the incident radiation field,
on the photodissociation and photoionization cross-sections, and on
the dust grain scattering and absorption properties. Dust attenutation
is the primary mechanism for the reduction of the photorates.
Line-shielding is also important
for the special cases of H$_2$ and CO photodissociation
\footnote{
We set the Doppler parameter equal to 2 km s$^{-1}$ for all 
absorption lines. As in SD95, we include H$_2$ transitions
in the Lyman and Werner bands, neglecting the rotational structure.
We use the Federman et al.~(1979) self-shielding formulae for H$_2$, and
the van Dishoeck \& Black (1989) shielding function for CO.
}.
For the depth dependent photodissociation and photoionization rates,
we adopt the biexponential representations 
\begin{equation}\label{atten}
\Gamma_i =  C_i{\rm exp}(-\alpha_i A_V - \beta_i A_V^2)
\end{equation}
and coefficients, $C_i$, $\alpha_i$, and $\beta_i$, 
calculated by SD95 for the radiative transfer of a Draine field penetrating 
a cloud of large ($A_{Vtot}=100$) total optical depth (Roberge et al.~1991).
We assume a dust-to-gas ratio such that 
$A_V=4.75\times 10^{-22}(N_{\rm H}+2N_{{\rm H_2}})$ 
(SD95; see also Draine 2003)
where $N_{\rm H}$ and $N_{{\rm H_2}}$ are the atomic and molecular
hydrogen column densities~(cm$^{-2}$). 
In Figure 1 we plot the CN and HCN
photodissociation rates as functions of $A_V$,
for $\chi=1$ and a total cloud thickness $A_{Vtot}=100$.
The much more rapid attenuation
of the CN photodissociation rate compared to HCN is an important
feature of the models, and reflects the fact that the energetic
photons ($> 12.4$ eV) required to dissociate CN (van Dishoeck 1987; 
Thi et al.~2004) are more readily absorbed by the dust grains. 

The cosmic-ray destruction rates $\xi_{ij}$ in equation (1)
include impact ionizations 
of H, H$_2$ and He, by
the primary cosmic-rays and secondary electrons, and induced FUV
photodissociation and photoionization of the heavier
molecules (Gredel et al.~1989).
In most of our models we adopt 
a ``dense cloud'' H$_2$ cosmic-ray ionization rate,
$\zeta = 5\times 10^{-17}$ s$^{-1}$
(e.g., Williams et al.~1998), but we also examine the
effect of increasing $\zeta$ by up to a factor 10
(McCall et al.~2003; Brittain et al.~2004).  
The cosmic-rays are ``freely penetrating'' and
$\zeta$ is independent of cloud depth. Cosmic-ray ionization
is the only source of H$^+$, H$_2^+$, and He$^+$ in our models.
X-ray ionization 
(Maloney et al.~1996; Spaans \& Meijerink 2005) is excluded.

The cosmic-ray induced photoionization and photodissociation 
rates are given by
\begin{equation}
\xi_i = \frac{\zeta p_i}{1-\omega}
\end{equation}
where $\omega$ is the grain albedo and $p_i$ are the photoabsorption
``efficiency factors'' (Gredel et al.~1989).
We adopt the efficiency factors $p_i$ listed in the 
{\small UMIST99} compilation, and we set $\omega = 0.5$ (SD95).

Molecule formation requires the presence of H$_2$. 
We assume that hydrogen molecules are formed on grain surfaces
with a rate coefficient 
\begin{equation}
R=3\times 10^{-18}T^{1/2}y_{\rm F} \ \ \ {\rm cm}^3 \ {\rm s}^{-1}
\end{equation}
where $T$ is the gas temperature, and $y_{\rm F}$ is a
``formation efficiency'' that depends primarily on the grain
temperature (Manicò et al.~2001; Biham et al.~2001, 2002; 
Cazaux \& Tielens 2002; Vidali et al.~2004). 
We set $y_F=1$ appropriate for low-temperature
($\sim 10-20$ K) dust grains. We do not consider
accretion or ejection of other molecular species onto, or from, the grains.

PDRs consist of several distinct zones or layers with sizes
and locations that depend on the gas density and FUV field strength.
As discussed by SD95, these include the outer H~{\small I} zone, 
followed by the C {\small II}, S {\small II}, and
Si {\small II} zones, terminated by a 
fully cosmic-ray dominated dark core. In the H {\small I} zone,
rapid photodissociation keeps the hydrogen in
atomic form. In the C {\small II} zone, 
the hydrogen ``self-shields'' and
becomes molecular, but photoionization maintains
the carbon as C$^+$. 
In the S {\small II}, zone the carbon
is fully incorporated into CO, but photoionization and charge-transfer
maintain the sulfur as S$^+$. 
In the Si {\small II} layer
the silicon is maintained as Si$^+$, but the sulfur is
either atomic 
or is incorporated into molecules. The relative efficiencies
of the various molecular formation sequences,
and the resulting molecular densities, vary with cloud
depth and from zone to zone.  For example, in the model 
considered by SD95, for which the hydrogen particle density
$n=10^6$~cm$^{-3}$ and $\chi=2\times 10^5$,  
the CN/HCN density ratio is large ($\sim 10$) in the
the H~{\small I}, C~{\small II}, and 
S~{\small II} layers, but then decreases and becomes small
($\sim 10^{-4}$) in the dark core.  A large CN/HCN ratio was
therefore identified as an important feature of (dense) PDRs.
SD95 identified additional molecular diagnostic ratios
of FUV irradiated gas,
including OH/H$_2$O, CO$^+$/HCO$^+$, and SO$^+$/SO
(see also Sternberg et al.~1996).

The sizes and locations of the various zones also depend
on the total gas-phase elemental abundances.
In our calculations we set the
gas-phase elemental abundances equal to the interstellar values observed in
the diffuse cloud toward $\zeta$ $Oph$. Thus, we set 
C/H=$1.32\times 10^{-4}$ (Cardelli et al.~1993), 
N/H=$7.50\times 10^{-5}$ (Meyer et al.~1997),
O/H=$2.84\times 10^{-4}$ (Meyer et al.~1998),
Si/H=$1.78\times 10^{-6}$ (Cardelli et al.~1994), and
S/H=$8.30\times 10^{-6}$ (Lepp et al.~1988). We
adopt a helium abundance He/H=0.1. In most of our models
we keep the gas-phase abundances
fixed at these values, and independent of cloud depth.
In one model we examine the effect of reducing the
S (and Si) abundances by a factor of 100.

In this paper, we study the behavior in cool isothermal clouds.
Our results are insensitive to the temperature
for $T\lesssim 200$ K,
and we adopt a representative gas temperature of $T=50$~K.
We consider isobaric clouds, so that the
total hydrogen particle densities $n\equiv n_{\rm H}+n_{\rm H_2}$ are
constant through the clouds. The densities of all other species
increase by a factor of 2 at the H to H$_2$ transition layer.  
We present results for $n$ ranging from 10$^2$ to 10$^6$ cm$^{-3}$.

Explicit thermal balance
computations (Sternberg \& Dalgarno 1989; Burton et al.~1990) 
show that in dense ($n\gtrsim 10^4$ cm$^{-3}$) PDRs
the temperatures in the outer H~{\small I} zones may exceed 10$^3$~K.
Molecular synthesis may then proceed via a ``hot gas'' chemistry
in a thin outer layer
(e.g.~via the efficient production of the reactive intermediate OH). 
However, the contributions to the total CN and HCN column
densities from the hot layers are expected to be small (SD95).
Here we exclude considerations of hot gas, and focus on the
chemistry 
in the more extended cooler regions.


\section{CN and HCN Formation and Destruction}

There are three major pathways to the gas-phase formation
of CN and HCN,
as illustrated in Figures 2-4. In pathway~\#1, production occurs
via the formation of carbon hydrides
and intermediate CH and CH$_2$ radicals. In pathway~\#2
an additional route to CN (but not HCN) occurs 
via oxygen hydrides and the intermediates OH and NO. 
In pathway~\#3 the molecules are produced
via nitrogen hydrides and parent H$_2$CN$^+$ ions.
We describe each of these pathways in turn.

\subsection{Pathway \#1}

In pathway \#1, which is important at all cloud depths, 
CN and HCN are formed via the neutral-neutral reactions
\footnote{
Reactions (R1) and (R2) play central roles in the formation 
of CN and HCN. In our computations we assume a rate coefficient of
$2.0\times 10^{-10}$ cm$^3$ s$^{-1}$ at 50 K for (R1),
as implied by pulsed laser experiments
in room-temperature discharge flows, and a theoretical
extrapolation to low temperature (Brownsword et al.~1996).
This is an order-of-magnitude larger than the rate coefficient 
reported by Messing et al.~(1981), and widely used in many
previous PDR models including SD95. For Reaction (R2) 
we use the theoretical value of $5.9\times 10^{-11}$
cm$^3$ s$^{-1}$ recently calculated by Herbst et al.~(2000).}
$$
{\rm CH + N \rightarrow CN + H}                 \eqno(R1)
$$
and
$$
{\rm CH_2 + N \rightarrow HCN + H} \ \ \ .      \eqno(R2)
$$

The required CH and CH$_2$ intermediates are produced in three ways.
First is via radiative association
$$
{\rm C^+ + H_2 \rightarrow CH_2^+ + \nu}        \eqno(R3)
$$
followed by rapid abstraction
$$
{\rm CH_2^+ + H_2 \rightarrow CH_3^+ + H}        \eqno(R4)
$$
and dissociative recombination
$$
{\rm CH_3^+ + e \rightarrow CH_2 + H}                           \eqno(R5)
$$
$$
{\rm CH_3^+ + e \rightarrow CH + H_2}         .                 \eqno(R6)
$$
Second is cosmic-ray driven proton transfer
$$
{\rm H_2 + cr \rightarrow H_2^+ + e}                               \eqno(R7)
$$
$$
{\rm H_2^+ + H_2 \rightarrow H_3^+ + H}                              \eqno(R8)
$$
$$
{\rm C + H_3^+  \rightarrow  CH^+ + H_2}                                \eqno(R9)
$$
followed by
$$
{\rm CH^+ + H_2 \rightarrow CH_2^+ + H}                              \eqno(R10)
$$
and then (R4), (R5), and (R6). 

A third possibility is direct formation 
\footnote{
Reaction (R11) was not included in the SD95
study. Here we adopt
$k_{11}=1.0\times 10^{-17}$ cm$^3$ s$^{-1}$
for the rate coefficient (Smith et al.~2004).
}
via slow radiative association
$$
{\rm C + H_2 \rightarrow CH_2 + \nu}                                 \eqno(R11)
$$
which may become competitive in dense clouds where the
fractional ionizations and the relative H$_3^+$ densities,
 $n_{\rm{H_3^+}}/n$, become small.

In the outer PDR, the CH and CH$_2$ radicals are removed by 
FUV photodissociation and photoionization.
In shielded regions they are removed by
$$
{\rm CH + O \rightarrow CO + H}                                    \eqno(R12)
$$
$$
{\rm CH_2 + O \rightarrow CO + H_2}                                \eqno(R13)
$$
in rapid reactions with the available oxygen atoms.

\subsection{Pathway \#2}

Pathway \#2 is an additional route to CN (but not HCN), and
becomes important mainly in shielded regions. In this pathway CN
is formed via the intermediates OH and NO in the reaction pair
$$
{\rm OH + N \rightarrow NO + H}                                    \eqno(R14)
$$
$$
{\rm NO + C \rightarrow CN + O} \ \ \ .                              \eqno(R15)
$$
The main source of OH is cosmic-ray driven proton transfer
$$
{\rm O  + H_3^+ \rightarrow OH^+ + H_2}                               \eqno(R16)
$$
followed by abstractions
$$
{\rm OH^+ + H_2 \rightarrow  H_2O^+ + H}                          \eqno(R17)
$$
$$
{\rm H_2O^+ + H_2 \rightarrow H_3O^+ + H}                        \eqno(R18)
$$
and dissociative recombination          
$$
{\rm H_3O^+ + e \rightarrow OH + H + H}                       \eqno(R19) 
$$
$$
{\rm H_3O^+ + e \rightarrow OH +H_2} \ \ \ .                   \eqno(R20)
$$
Major removal mechanisms for OH and NO are
photodissociation, and reactions with atomic
oxygen and nitrogen
$$
{\rm OH + O \rightarrow O_2 + H}                             \eqno(R21)
$$
$$
{\rm NO + N \rightarrow N_2 + O}                              \eqno(R22)
$$
in the shielded regions.

\subsection{Pathway \#3}

Finally, in pathway \#3 both CN and HCN are
produced via dissociative recombination
$$
{\rm H_2CN^+ + e \rightarrow HCN + H}                          \eqno(R23)
$$
$$
{\rm H_2CN^+ + e \rightarrow CN + H_2}                        \eqno(R24)
$$
where the parent H$_2$CN$^+$ ions are formed via 
$$
{\rm NH + C^+ \rightarrow CN^+ + H}                            \eqno(R25)
$$
$$
{\rm NH_2 + C^+ \rightarrow HCN^+ + H}                       \eqno(R26)
$$
followed by
$$
{\rm CN^+ + H_2 \rightarrow HCN^+ + H}                      \eqno(R27)
$$
$$
{\rm HCN^+ + H_2 \rightarrow H_2CN^+ + H} \ \ \ .            \eqno(R28)
$$
This pathway is mainly important in shielded regions,
but may also operate in the outer H~{\small I} zone.

In shielded regions, where the N$_2$ density
becomes large, NH and NH$_2$ are
produced by the cosmic-ray driven sequence
$$
{\rm He^+ + N_2 \rightarrow N^+ + N + He}                  \eqno(R29)
$$
$$
{\rm N^+ + H_2 \rightarrow NH^+ + H}                               \eqno(R30)
$$
$$
{\rm NH^+ + H_2 \rightarrow NH_2^+ + H}                             \eqno(R31)
$$
$$
{\rm NH_2^+ + H_2 \rightarrow NH_3^+ + H}                            \eqno(R32)
$$
followed by
$$
{\rm NH_3^+ + e \rightarrow NH + H_2}                               \eqno(R33)
$$
$$
{\rm NH_3^+ + e \rightarrow NH_2 + H}                               \eqno(R34)
$$
In the H~{\small I} zone, NH may be formed via
$$
{\rm N + H_2^* \rightarrow NH + H}                                \eqno(R35)
$$
where H$_2^*$ refers to the sum over all
FUV-pumped hydrogen molecules in vibrational states
with energies large enough to overcome the
endothermicity of the reaction.

\subsection{Destruction}

The CN and HCN density profiles also depend on the varying efficiencies 
of the destruction mechanisms.
In the outer PDR, photodissociation
$$
{\rm CN + \nu \rightarrow C + N}                               \eqno(R36)
$$
$$
{\rm HCN + \nu \rightarrow CN + H}                             \eqno(R37)
$$
dominates. 
In shielded layers the CN radical is removed by
$$
{\rm CN + N \rightarrow N_2 + C}                            \eqno(R38)
$$
$$
{\rm CN + O \rightarrow CO + N} \ \ \  .                       \eqno(R39)
$$
Reaction (R38) dominates
at intermediate cloud depths where a high abundance of nitrogen
atoms are maintained by photodissociation. At large depths
(R39) becomes the dominant removal process. 
The saturated HCN molecule
is removed primarily by cosmic-ray induced photodissociation
$$
{\rm HCN + crp \rightarrow CN + H} \ \ \ .                       \eqno(R40)
$$
Photodissociation of HCN operates as an important source of CN,
both in the PDR and in the dark core.  
Additional removal processes for HCN in the dark core are
$$
{\rm HCN + H^+ \rightarrow HCN^+ + H}                             \eqno(R41)
$$
$$
{\rm HCN + HCO^+ \rightarrow H_2CN^+ + CO}                        \eqno(R42)
$$
Reaction (R41) is important in low density clouds 
($n \lesssim 10^3$ cm$^{-3}$)
where a large proton density is maintained by cosmic-ray
ionization. Reaction (R42) becomes important in dense clouds 
($n \gtrsim 10^3$ cm$^{-3}$).

\subsection{Free Carbon: C$^+$ and C}

The efficiencies of the CN and HCN formation sequences
depend on the availability of ``free carbon'' particles, either as 
C$^+$ ions or C atoms. As we discuss in \S 4, 
the CN and HCN profiles are closely related to the
C$^+$ and C density distributions. 

In the outer parts of the PDR,
C$^+$ is produced by FUV photoionization
$$
{\rm C + \nu \rightarrow C^+ + e} \ \ \ .                      \eqno(R43)
$$
In shielded regions the main source is
cosmic-ray driven helium-impact dissociative ionization of CO
$$
{\rm He + cr \rightarrow He^+ + e}                    \eqno(R44)
$$
$$
{\rm He^+ + CO \rightarrow C^+ + O + He} \ \ \ .         \eqno(R45)
$$
In the outer PDR, where the electron density is high,
the C$^+$ ions are removed by radiative recombination
$$
{\rm C^+ + e \rightarrow C + \nu} \ \ \ .                          \eqno(R46)
$$
In shielded regions, radiative association (R3), and 
charge transfer neutralization
$$
{\rm C^+ + S \rightarrow S^+ + C} \ \ \ ,                  \eqno(R47)
$$
become important. 
Charge transfer with atomic sulfur plays a particularly
important role in the partially shielded layers.
In the dark core
$$
{\rm C^+ + O_2 \rightarrow CO + O^+}                         \eqno(R48)
$$
$$
{\rm C^+ + O_2 \rightarrow CO^+ + O}                       \eqno(R49)
$$
dominate.

Carbon atoms are produced by radiative recombination and 
charge transfer ([R46] and [R47]), FUV photodissociation
$$
{\rm CO + \nu \rightarrow C + O}     \eqno(R50)
$$
and cosmic-ray induced photodissociation
$$
{\rm CO + crp \rightarrow C + O} \ \ \ .                           \eqno(R51)
$$
The C atoms are removed by 
photoionization in the outer PDR, by proton transfer 
and radiative association ([R9] and [R11]), and 
cosmic-ray induced photoionization
$$
{\rm C + crp \rightarrow C^+ + e}                  \eqno(R52)         
$$
at intermediate depths, and by
$$
{\rm C + O_2 \rightarrow CO + O}           \eqno(R53)
$$
$$
{\rm C + SO \rightarrow S + CO}            \eqno(R54)
$$
$$
{\rm C + SO \rightarrow CS + O}            \eqno(R55)
$$
in the shielded cores.


\section{Reference Model}

The results of our model calculations are displayed in
Figures 5, 6, and 7, which show cuts through
the $\chi$, $n$, $\zeta$ parameter space we have considered.
For each set of cloud parameters we display the 
C$^+$, C, CO, CN, and HCN densities (cm$^{-3}$) and
column densities (cm$^{-2}$), as well as the
CN/HCN density and column density ratios, as functions of $A_V$.
The column densities are integrated from the cloud surface,
in the normal direction.  We also display the electron density profiles.

We first discuss our results for a specific
``reference model'', with 
$n=10^4$~cm$^{-3}$, $\chi=2\times 10^3$,
 and $\zeta=5\times 10^{-17}$ s$^{-1}$.
We use this model to analyze the depth dependent  
formation sequences and resulting density profiles. We also use it as
a point of comparison as the cloud parameters are varied.
Our reference model is displayed in panels (c), (d), and (e) of Figure 5,
and again in Figures 6 and 7. 

In our reference model, and in our parameter study we adopt $T=50$ K.
For our assumed set of reaction rate coefficients,
the results we describe below are quite insensitive to the assumed gas
temperature, provided the gas is cold with $T\lesssim 200$ K. 
For example, relative to our results for 50 K, the CN and HCN densities 
change by less than a factor of 2 at all $A_{\rm V}$,
if $T$ is decreased to 20 K, or increased to 200 K.

\subsection{C$^+$, C, and CO Density Profiles}

The C$^+$, C, and CO density profiles in our reference model show the
familiar structure seen in numerous PDR calculations 
(Tielens \& Hollenbach 1985; van Dishoeck \& Black 1988;
Le Bourlot et al.~1993a; Flower et al.~1994; SD95; Jansen et al.~1995;
Bakes \& Tielens 1998).
The carbon is ionized in the outer layers, and undergoes a conversion to CO as
the ionizing FUV radiation is attenuated. An important feature is the
double peaked atomic carbon density profile.

The first C peak is associated with the sharp transition
of the available gas-phase carbon from fully ionized to molecular form
near the inner edge of the C~{\small II} zone. In our reference
model this occurs at $A_V=2$. 
Due to the rapid production of CO at this location, the C density at
the peak reaches only $\sim 20\%$ of the total gas-phase carbon abundance.
The transition is, effectively, directly from C$^+$ to CO.
After the conversion to CO only trace quantities of C and C$^+$ remain. 

An important property of 
the C$^+$/C/CO transition region is that photodissociation
of CO continues to dominate the production of the free carbon
well beyond the point where the conversion to CO is complete.
For example, in
our reference model, photodissociation dominates up to 
$A_V=2.5$, well inside the S {\small II} zone. Up to this point,
C$^+$ continues to be produced by photoionization, and the 
C$^+$/C density ratio is
set by the balance between recombination and photoionization.
Thus, in the C {\small II} zone, where the production rate of
atomic carbon via recombination is constant 
(since $n_{\rm e}\approx n_{\rm C^+}$),
the C density increases as the photoionization
rate is attenuated. However, once the conversion to CO is complete
the atomic carbon density drops as the 
CO photodissociation rate decreases. 
This behavior gives rise to the first C peak.

We note that for PAH
abundances $\gtrsim 10^{-7}$ mutual neutralization,
${\rm C}^+ + {\rm PAH}^- \rightarrow {\rm C + PAH}$,
can compete with radiative recombination (Lepp et al.~1988)
leading to a slight
shift in the positions of the C$^+$/C/CO transition layers
(Bakes \& Tielens 1998).
We do not include the effects of PAHs in our analysis.
Similarly, the process of ``grain-assisted'' 
ion-electron recombination
may become important when
$n_{\rm e}\alpha_r/n\alpha_g < 1$, where $\alpha_r$ 
(equal to $1.4\times 10^{-11}$ cm$^{-3}$ s$^{-1}$ at 50 K, for
C$^+$ ions), 
and $\alpha_g\lesssim 10^{-14}$ cm$^3$ s$^{-1}$
are the radiative and grain-assisted recombination rate
coefficients.  The efficiency of grain-assisted recombination
depends on the grain population size distribution and on the
grain charge (Weingartner \& Draine 2001). In the computations
we present here we exclude this process. Including it
can increase the C densities at the first C peak.
We have verified that the CN and HCN density profiles
we discuss below are not affected significantly.

At larger depths, e.g. for $A_V > 2.5$ in our model,
the rate of CO photodissociation becomes vanishingly small, and the 
primary source of free carbon becomes
cosmic-ray driven helium-impact ionization of CO
([R44] and [R45]). The resulting C$^+$ ions
are neutralized by radiative recombination (R46),
and by rapid charge transfer of the C$^+$ ions
with S atoms~(R47).
The rapid neutralization keeps the free carbon in atomic form,
so that $n_{\rm C}/n_{\rm C^+} \gg 1$. 
Because the rate of helium impact ionization
is independent of cloud depth, the C
density {\it increases} with $A_V$ so long as photoionization 
dominates the removal of the carbon atoms,
and as the photoionization rate is attenuated.
The second C peak occurs at the point where 
radiative association with H$_2$ (R11),
cosmic-ray induced photoionization (R52), and
proton transfer reactions with H$_3^+$ (R9) begin
to dominate the removal of the carbon atoms. In our reference model
this occurs at $A_V=3.7$. 

The C density at this location may be estimated analytically,
\begin{equation}\label{cden}
n_{\rm C}\approx 
\frac{\zeta X_{\rm He}}{2p_{52}\zeta/n + k_9n_{\rm {H_3^+}}/n + k_{11}}
\end{equation}
where $k_9=2.0\times 10^9$ and $k_{11}=1.0\times 10^{-17}$
cm$^3$ s$^{-1}$ are the rate coefficients for (R9) and (R11),
the parameter
$p_{52}=1.0\times 10^{3}$ is the efficiency factor for (R52), and
$X_{\rm He}=0.1$ is the He abundance. The first two terms in the
denominator becomes small for $n \gtrsim 10^4$ cm$^{-3}$,
and $n_{\rm C}$ is then independent of $n$.
In this limit,
$n_{\rm C}\approx 0.25$ cm$^{-3}$ for $\zeta=5\times 10^{-17}$ s$^{-1}$,
in good agreement with the numerical
value of 0.40 cm$^{-3}$ at the second C peak.  
Beyond this point the C density decreases,
particularly as the O$_2$ and SO densities rise
and as the removal reactions
(R53), (R54) and (R55) become more effective. 
A sharp drop in the C density occurs at $A_V\approx 7$, and
the C density
finally stabilizes to a value of $3.6\times 10^{-4}$ cm$^{-3}$ in
the cosmic-ray dominated dark core.  The atomic carbon density in the
core depends on $n$ and $\zeta$ as discussed further in \S 5.

The second C peak 
marks the point where removal of the C atoms by FUV 
photoionization becomes
inefficient compared to chemical reactions 
and cosmic-ray induced photoionization.
The minimum between the two atomic carbon peaks occurs near to the point 
where helium impact dissociative ionization of CO replaces 
photodissociation of CO as the primary source of free carbon particles. 

\subsection{CN and HCN Profiles}

In our reference model a
pronounced peak is present in the CN density profile 
at $A_V = 2$, at the inner edge of the C {\small II} 
zone, and close to the location of the first C peak.
A small HCN peak is also present at this location.
Throughout the C {\small II} zone, the CN and HCN molecules
are produced primarily via
pathway \#1, where the formation of 
the CH and CH$_2$ intermediates is initiated
by radiative association of C$^+$ with H$_2$ (R3). The CN density peak occurs
where the C$^+$ density is still close to its maximum possible value,
but where the photodissociation rates are diminished.
The CN density then decreases with cloud depth as
the C$^+$ density drops across the C$^+$/C/CO transition layer.

Because the carbon is fully ionized in the C {\small II} zone,
molecule formation is ``recombination limited'' in this part
of the cloud. 
The CN density at the inner edge of the C {\small II} zone
may therefore be expected to scale linearly with the gas 
density $n$ for a fixed value of the ``ionization parameter'' $\chi/n$.
This may be seen analytically by writing
\begin{equation}\label{CNpeak}
n_{\rm CH} \approx \frac{k_3 n_{{\rm C}^+} n_{{\rm H}_2}}{\chi \Gamma_{\rm CH}^\prime}f_{\rm CH}
\approx \frac{k_3}{\chi \Gamma_{\rm CH}^\prime}f_{\rm CH}X_{\rm C}n^2
\end{equation}
and
\begin{equation}\label{CNscale1}
n_{\rm CN} \approx \frac{k_1 n_{\rm CH}n_{\rm N}}{\chi \Gamma_{\rm CN}^\prime}
\approx
\frac{k_1 k_3}
{\Gamma_{\rm CH}^\prime \Gamma_{\rm CN}^\prime}f_{\rm CH}X_{\rm C} X_{\rm N}
\biggl(\frac{n}{\chi}\biggr)^2n \ \ \ .
\end{equation}
In these expressions, 
$k_1=2.0\times 10^{-10}$ and $k_3=5.7\times 10^{-16}$ cm$^3$ s$^{-1}$
are the rate coefficients
for reactions (R1) and (R3), and 
$\chi\Gamma_{\rm CH}^\prime=1.7\times 10^{-8}$ and 
$\chi\Gamma_{\rm CN}^\prime=5.0\times 10^{-10}$ s$^{-1}$
are the {\it attenuated} photodissociation rates at the
inner edge of the C {\small II} zone.
The fraction of recombining of CH$_3^+$ ions that fragment to CH
is $f_{\rm CH}=0.43$, and
$X_{\rm C}=1.32\times 10^{-4}$ and
$X_{\rm N}=7.50\times 10^{-5}$ are the 
gas phase abundances of carbon and nitrogen. 
For $n=10^4$ cm$^{-3}$ and $\chi = 10^3$
this gives $n_{\rm CN}=5.7\times 10^{-5}$ cm$^{-3}$,
in good agreement with our numerically computed 
value of $9.0\times 10^{-5}$ cm$^{-3}$ at the peak.
The photo-attenuation factors (see equation [\ref{atten}])
depend on the visual extinctions
at the termination points of the C {\small II} zones, and therefore scale with $\chi/n$. 
Equation (\ref{CNscale1}) shows that for fixed $\chi/n$ the CN density at the peak is 
proportional to the product of the carbon and nitrogen abundances and
to the cloud density $n$, and is independent of $\chi$.
These simple scalings are verified by our numerical results in \S 5.
 
The CN density decreases as the
supply of free carbon produced by CO photodissociation diminishes.
However, when the primary source of free carbon switches to helium impact
ionization, the CN density stops decreasing and
a pronounced ``plateau'' appears in the density profile. In our
reference model the CN density plateau extends from $A_V=3$ to 6.

The CN density profile flattens for a combination of reasons.
First, while the CH and CH$_2$ formation efficiencies
decrease as the C$^+$ disappears,
this is offset by a more rapid decline in the
the CH and CH$_2$ photodestruction rates.
Second, the {\it rise} in the atomic carbon
density up to the second C peak leads to 
an increased efficiency of the proton-transfer
sequence initiated by (R9) in pathway \#1. Third, CN formation via
pathway \#2 also begins to play a significant role
as the OH and NO densities increase. 

Furthermore, while HCN (formed via [R2]) is removed by photodissociation, 
at these cloud depths
CN is removed by neutral-neutral reactions
with N atoms (R38). Photodestruction of CN is ineffective in the plateau
because of the severe attenuation of the CN photodissociation rate (see Figure 1).
Importantly, photodissociation of HCN becomes a major source of CN in the plateau.
The HCN density therefore rises as the FUV field is attenuated,
but because the atomic nitrogen density remains large throughout,
the resulting CN density remains insensitive to $A_V$. 

At these cloud depths molecule formation is ``ionization limited'' since
only a trace amount of free-carbon is released from the
CO molecules by helium impact ionization.
Because CN is removed
in reactions with a neutral species (N~atoms), the
CN density is proportional to the cosmic-ray ionization rate
and independent of the gas density.  For a simple analytic estimate we
assume that CN formation is initiated by proton transfer in
reactions of C with H$_3^+$ (R9), or by
radiative association of C with H$_2$ (R11),
in pathway \#1. We also assume
and that every HCN formation event leads
to CN via photodissociation. We assume further that
the CH and CH$_2$ intermediates are removed by reactions
with atomic oxygen [(R12) and (R13)], and that these reactions proceed 
with equal approximate rate coefficients 
$k_{\rm O}=2\times 10^{-10}$ cm$^3$ s$^{-1}$.
Setting $k_{\rm N}=1\times 10^{-10}$ cm$^3$ s$^{-1}$
for both (R1) and (R2) it follows
\footnote{The factor of 0.5 in this expression enters because the
cosmic ray ionization rate of He is half that of H$_2$.} that
\begin{equation}\label{CNplat}
n_{\rm CN} \approx 0.5 \zeta \frac{1}{k_{38}}\frac{k_{\rm N}}{k_{\rm O}}
\frac{X_{\rm He}}{X_{\rm O}}
\end{equation}
where $k_{38}=3\times 10^{-10}$ cm$^3$ s$^{-1}$
is the rate coefficient for removal reactions of CN with N. 
Evaluating for $X_{\rm He}=0.1$, $X_{\rm O}=2.84\times 10^{-4}$,
and $\zeta=5\times 10^{-17}$ s$^{-1}$ yields 
$n_{\rm CN}=9\times 10^{-6}$ cm$^{-3}$, in good agreement
with the numerically computed
$n_{\rm CN}\approx 2\times 10^{-5}$ cm$^{-3}$ 
in the CN plateau (see Figure 5).
Thus, $n_{\rm CN}$ is proportional to $\zeta$ and independent of $n$,
and with the above assumptions
\footnote{These assumptions break down if, for example,
$X_C > X_O$ so that C atoms become abundant rather than
being mainly locked in CO, or if $X_C$ becomes so small
such that formation via pathway \#1 becomes negligible.
},
is independent of both the gas-phase nitrogen and
carbon abundances.

The CN density plateau is maintained up to the point where
photodissociation by the incident FUV photons
is the dominant HCN removal mechanism, or up to $A_V=6$
in our reference model. The HCN density reaches a
peak value of $8.6\times 10^{-4}$ cm$^{-3}$ at this point.
At larger depths cosmic-ray induced photodissociation 
becomes the dominant HCN destruction mechanism, 
proceeding at a rate independent of $A_V$.
CN is removed by reactions with oxygen atoms (R39).
The CN and HCN formation efficiencies both decline
as less C and C$^+$ is available,
and the CN and HCN densities decrease to their 
dark core values of $1.2\times 10^{-6}$ and 
$2.6\times 10^{-5}$ cm$^{-3}$ respectively.

Figure 5e shows that the CN column density rise sharply
to $N_{\rm CN} = 6\times 10^{12}$ cm$^{-2}$ at
the CN peak at $A_V=2$, and that much of the
CN column is built up at this location. The HCN
column remains small at $A_V=2$, but
continues to increase with cloud depth. $N_{\rm HCN}$ 
approaches  $7\times 10^{14}$ cm$^{-2}$ at $A_V=6$
where the HCN volume density is at its maximum.
The ratio $N_{\rm CN}/N_{\rm HCN}\sim 10$ at $A_V=2$,
and decreases to $\sim 0.1$ at $A_V=6$ (see Figure~5f).


\section{Parameter Study}

 
\subsection{Density}

Figure 5 displays our results for $n$ equal to
$10^3$, 10$^5$, and 10$^6$ cm$^{-3}$, in addition to
our $10^4$ cm$^{-3}$ reference model, 
and illustrates the effects
of varying the cloud density. In this sequence we keep the ionization
parameter $\chi/n$ constant at 0.2 cm$^{-3}$, so that
the C$^+$/C/CO transition layers occur at
the same location ($A_V=2$) in all four models.

For fixed $\chi/n$, the depth at which helium impact ionization of CO
replaces photodissociation of CO as the source of free carbon,
increases with the cloud density $n$.
However, the depth at which removal of atomic carbon by
photoionization becomes ineffective compared to proton transfer or
radiative association ([R9] or [R11]), is independent of $n$.
Therefore, the relative height of the second C peak (at $A_V=4.2$)
decreases with $n$, and the peak disappears at sufficiently high densities
($\gtrsim 10^5$ cm$^{-3}$). For densities above $10^4$ cm$^{-3}$,
$n_{\rm C}\approx 0.4$ cm$^{-3}$ at this location,
consistent with Equation~(\ref{cden}).
At lower densities all of the
available carbon is atomic at the location of the second C peak,
and $n_{\rm C}$ is then limited by the carbon abundance $X_{\rm C}$ 
(see Figure 5a).

At low $n$, the free carbon density remains 
relatively high at large $A_V$ and in the dark core.
At high $n$ the free carbon density becomes small.  
As we discuss below, this behavior reflects the
transition from the ``high-ionization phase'' to 
``low-ionization phase'' in the cosmic-ray dominated core 
(Le Bourlot et al.~1993b; 1995; Lee et al.~2000).
The asymptotic values of the C densities in the dark core
influence the shapes of the atomic carbon density
profiles at intermediate depths in the PDRs.   

Figure 5 shows that the CN density at $A_V=2$ increases
linearly with $n$,
from $\sim 10^{-5}$ to $10^{-2}$ cm$^{-3}$,
in accordance with Equation~(\ref{CNpeak}). The ``CN peak''
becomes more pronounced as $n$ increases. In the 
$n=10^6$ cm$^{-3}$ model, the density of FUV-pumped H$_2$ 
becomes large, and CN is also formed via pathway \#3
initiated by (R35). 
For $n\ge 10^4$ cm$^{-3}$ the ``CN plateau'' from
$A_V=3$ to $\sim 6$ is apparent. At these depths
$n_{\rm CN}\approx 2\times 10^{-5}$ cm$^{-3}$
independent of $n$, consistent with Equation~(\ref{CNplat}). 
At low $n$, the CN density remains large to
high $A_V$ due to the elevated free carbon densities.

For fixed $\chi/n$, FUV photodissociation
of HCN is effective to greater
depths as $n$ increases. The location of the
HCN density peak therefore moves from a visual
extinction of $5.6$ to $7.2$ as the gas density is increased
from $10^4$ to $10^6$ cm$^{-3}$.
The column density ratio $N_{\rm CN}/N_{\rm HCN}$ remains
large to greater depths as $n$ is increased,
but the density ratio $n_{\rm CN}/n_{\rm HCN}$ falls
off sharply with $A_V$ (see Figure 5).

\subsection{FUV Intensity, Sulfur Abundance, and Cosmic-Ray Ionization Rate}

In Figure 6, we set $n$ equal to
$10^4$ cm$^{-3}$, and we vary $\chi$ from $2\times 10^2$
to $2\times 10^6$. This illustrates the effects of varying the
incident FUV field intensity.
The location of the C$^+$/C/CO transition layer moves
from $A_V=1.3$ to 3.5 for this density and range
of FUV intensities. The position of the second C peak
is also shifted to larger $A_V$,
as is the location of the sharp drop in the C density,
which moves from $A_V=6$ to 9. 
The approach to the dark-core conditions occurs at greater
$A_V$ as $\chi$ is increased.

Correspondingly, the locations of the CN peak and the
inner HCN peak also move inward as $\chi$ is increased.
Because the CN photodissociation rate declines most rapidly
with $A_V$ (see Figure 1) 
$n_{\rm CN}/n_{\rm HCN}$ and $N_{\rm CN}/N_{\rm HCN}$ 
at the CN peak increase with $\chi$,
from $\sim 10$ to 40 in our models. 

In Figure 7, we show the effects of a reduced gas-phase
S (and Si) abundance, and an increased cosmic-ray ionization
rate, for our $n=10^4$ cm$^{-3}$ and $\chi=10^3$ model. In
panels (a), (b), and (c) of
Figure 7, the S (and Si) abundances are reduced by a factor of 100.
Charge transfer (R47) between C$^+$ and S is then less
effective as a neutralizing mechanism, and the second 
C peak at large $A_V$ is reduced.
The formation efficiency of HCN in pathway~\#1, via radiative
association of C with H$_2$ (R11) followed by
the reaction of CH$_2$ with N (R2),
is therefore also diminished, leading to a reduction
of the inner HCN density peak (compare Fig.~7a and Fig.~7d).
The resulting CN/HCN ratios are therefore enhanced in
the CN plateau region, and in the dark core.

The bottom two rows of Figure 7 show the effects of
increasing the cosmic-ray ionization rate, by factors of 3 and 10,
to $1.5\times 10^{-16}$ and $5\times 10^{-16}$ s$^{-1}$.
Increasing $\zeta$ mainly affects the atomic carbon density profile.
The first C peak is unaltered because it is 
is controlled by purely photo-processes.
However, the second C peak is enhanced by the increased rate
of helium impact destruction of CO (R44). The behavior 
is consistent with Equation~(\ref{cden}). Initially, the C density increases
linearly with $\zeta$. However, for sufficiently large
$\zeta$ the atomic carbon is removed by 
cosmic-ray induced photoionization (R52) and proton transfer
reactions with H$_3^+$ (R9), rather than by radiative association
with H$_2$ (R11). In this limit the C density is independent
of $\zeta$, and all of the available gas phase carbon becomes atomic,
as occurs for our $\zeta=5\times 10^{-16}$ model.

The free carbon densities remain large
throughout the cloud for high $\zeta$.
As we now discuss, the shift from a low to high free carbon
density with increasing $\zeta$ is related to
a ``phase change'' that occurs in the dark core.

\subsection{Dark Core}

In Figure 8 we display the densities, $n_i$, and density fractions, $n_i/n$,
for C$^+$, C, CO, CN, HCN, and free electrons,
for fully opaque dark core conditions. Externally incident
FUV photons are excluded, and the chemistry is
driven entirely by cosmic-ray ionization. 
For these conditions, it follows from Equation (1) that the
density fractions $n_i/n$ depend on a single parameter,
the ratio of the cloud density to the ionization rate.
In Figure 8 we plot the solutions as functions of
$n/\zeta_{-17}$, where $\zeta_{-17}$ is the 
cosmic-ray ionization rate
normalized to $1.0\times 10^{-17}$ s$^{-1}$. 
The phenomenon of ``bistability'' 
(Le Bourlot et al.~1993b, 1995; Lee et al.~2000)
is apparent in Figure 8, where for densities between 
370 and 675~cm$^{-3}$ two stable solutions exist.  
A high-ionization phase (HIP) occurs at low $n$, and
a low-ionization phase (LIP) occurs at high $n$.
The two phases may coexist where the gas is bistable.
We will present our own discussion of bistability elsewhere
(Boger \& Sternberg 2005). 
Because bistability occurs for a narrow range of densities
the transition from the HIP to LIP may be said to occur
near a ``critical density'' $n_{\rm crit}/\zeta_{-17} \approx 500$ cm$^{-3}$
(for our assumed gas-phase abundances).
In our computations, $n_e/n \gtrsim 10^{-5}$ in the HIP, and
$n_e/n \lesssim 10^{-6}$ in the LIP, and
the fractional ionization drops by a factor $\sim 10$,
at the transition density.
The C$^+$ and C densities 
are large in the HIP, with $n_{\rm C}/n_{\rm CO} \sim 1$,
and small in the LIP, with $n_{\rm C}/n_{\rm CO} < 10^{-3}$,
consistent with the findings of Flower et al.~(1994).
Correspondingly,
$n_{\rm CN}/n$ and $n_{\rm HCN}/n$ are large for $n<n_{\rm crit}$, and
small for $n>n_{\rm crit}$. Furthermore, $n_{\rm CN}/n_{\rm HCN}$ is 
large in the low-density HIP, and becomes small in the high-density LIP.
In the HIP, $n_{\rm CN}/n_{\rm HCN}$ decreases from $\sim 20$ to 
$0.07$ for $n/\zeta_{-17}$ ranging from 10 to 675 cm$^{-3}$.
In the LIP, $n_{\rm CN}/n_{\rm HCN}$ decreases
from 0.4 to $3\times 10^{-3}$ cm$^{-3}$ for $n/\zeta_{-17}$
ranging from 370 to 10$^5$ cm$^{-3}$.

Returning now to Figure 7, we note that 
the increase in the free carbon density at large $A_V$ that occurs
as $\zeta$ is increased, is due to the rise in $n_{\rm crit}$,
and a corresponding transition from LIP to HIP conditions, for the specific
gas density of $10^4$ cm$^{-3}$. 
The phase transition
is affected by the presence of some FUV radiation. For example,  
for $\zeta=1.5\times 10^{-16}$ s$^{-1}$ it occurs
at $A_V\approx 7$ rather than in the fully
opaque core (see Figure 7g). As the ionization rate is increased further,
the gas is converted to the HIP throughout (see Figure 7j).
 

\section{Discussion and Summary}

CN and HCN molecules have been observed in a diversity of Galactic and
extragalactic sources, and the CN/HCN intensity and density ratios have
been used as diagnostic probes of FUV irradiated molecular gas.
One recent and interesting example is
the well-known starburst galaxy M82
($d=3.9$~Mpc; $L=3.7\times 10^{10}$ L$_\odot$;
e.g.~F\"orster-Schreiber~et al.~2003).
Fuente~et~al.~(2005) have reported measurements
of CN~1-0~(113.490~GHz), CN~2-1~(226.874~GHz), and HCN~1-0~(88.631~GHz)
line emissions across the inner 650 pc star-forming molecular 
disk in M82. They find that $N_{\rm CN}/N_{\rm HCN} \sim 5$,
and argue that the large ratio is indicative of 
a giant and dense PDR bathed in the intense field of the starburst.
They derive $n \sim 10^4$ to $10^5$ cm$^{-3}$, 
and $\chi\sim 10^4$. Fuente et al.~(2005)
conclude that $A_V\lesssim 5$ in the M82 clouds, since for optically
thicker clouds the CN/HCN column density ratio would be smaller
than observed. Our results provide support for these conclusions.
In this picture $\sim 10$ to 20 individual clouds along the line
of sight are required, since for characteristic elemental abundances 
the computed CN and HCN columns 
for $A_V\lesssim 5$ (see Figures 5 and 6) are about an order of 
magnitude smaller than observed 
($N_{\rm CN} = [2 \pm 0.5] \times 10^{14}$~cm$^{-2}$ and
$N_{\rm HCN} = [4 \pm 0.5] \times 10^{13}$~cm$^{-2}$).

Alternatively, the large CN/HCN ratio might be a signature
of a very large cosmic-ray ionization rate, up to 
$\sim 5\times 10^{-15}$ s$^{-1}$, as invoked by Suchkov et al.~(1993)
for M82 and other starburst galaxies.
For such high ionization rates the CN/HCN
ratio could remain $\gtrsim 1$ even in dense and opaque cores
if these are maintained in the high-ionization phase.
Our results indicate that for $\zeta = 5\times 10^{-15}$ s$^{-1}$, the HIP
is maintained up to $n\sim 4\times 10^5$~cm$^{-3}$ (see Figure~8).
This possibility is perhaps more compatible with the large 
C/CO ratio $\gtrsim 0.5$ inferred from observations
of the 492 and 809 GHz C~{\small I} fine-structure lines in M82 
(Schilke~et~al.~1993; Stutzki~et~al.~1997; see also Gerin \& Phillips 2000).
As discussed above, a large C/CO ratio is a signature of the HIP.

In this paper we have presented a theoretical study 
of CN and HCN molecule formation in dense interstellar clouds
exposed to intense FUV radiation fields.
We have analyzed the behavior of the CN/HCN
density ratio for a wide range of conditions, with the aim of showing
how this molecular ratio may be used as
a diagnostic probe of molecule formation
in FUV irradiated gas.
For this purpose, we have constructed detailed models in which
we solve the equations of chemical equilibrium
as functions of optical depth, for uniform density clouds
at constant gas temperature (50~K). 
Our results are insensitive to the gas temperature for
cold clouds with $T\lesssim 200$ K.
We consider clouds with hydrogen particle densities
ranging from $n=10^2$ to $10^6$ cm$^{-3}$, and FUV radiation
intensities ranging from $\chi=20$ to $2\times 10^5$, appropriate for
star-forming clouds near young OB stars and clusters.
We present results for cosmic-ray ionization rates ranging from
$5\times 10^{-17}$ to $5\times 10^{-16}$ s$^{-1}$, 
and we also examine the effects of large (factor 100)
sulfur depletions on the computed density profiles.
We present calculations of the density profiles for 
CN and HCN, and for the associated species C$^+$, C, and CO.
We analyze the behavior from the outer FUV photon-dominated
regions, into the fully opaque cosmic-ray dominated cores.

In this paper, we adopt a fixed characteristic shape for
the FUV radiation spectrum, and we do not examine the possible
signatures of ``soft'' versus ``hard'' FUV fields 
(van Zadelhoff et al.~2003). Time-dependent effects due to 
episodic shadowing (St\"orzer et al.~1997), fluctuating
radiation fields (Parravano et al.~2003),
grain processing (Charnley et al.~2001), 
turbulent mixing and diffusion (Papadopoulos et al.~2004),
and other dynamical processes, are not considered here.

Our models show how observations of the CN/HCN
abundances ratio in molecular clouds may used as probes
of FUV and cosmic-ray driven gas-phase chemistry for
a wide range of conditions.
We find that in dense gas, CN molecules are characteristically and 
preferentially produced near the inner edges of the C~{\small II} zones
in the PDRs. This is where  
C$^+$ begins to recombine and where atomic carbon is
incorporated into CO. Molecule formation is
``recombination limited'' at these depths, and
for fixed $\chi/n$, the CN density
is proportional to the cloud density, and to the gas-phase carbon and nitrogen
abundances. For $n\gtrsim 10^4$~cm$^{-3}$, and for clouds 
with linear sizes corresponding
to visual extinctions $A_V\lesssim 10$, the entire 
integrated CN column density is built up at the C$^+$/C/CO transition layer. 
For characteristic interstellar 
carbon and nitrogen gas-phase abundances the 
predicted CN columns are $\sim 3\times 10^{13}$ cm$^{-2}$.
HCN is rapidly photodissociated in the outer parts of the PDRs
including the C$^+$/C/CO transition layers. Because HCN is more
vulnerable to photodissociation, the CN/HCN density ratio is large
at low $A_V$, and decreases with increasing optical depth.
We find that the CN/HCN density ratio typically decreases
from $\gtrsim 10$ in the C$^+$/C/CO transition layers
to $\lesssim 0.1$ in the opaque cores.

At intermediate and large depths, the bulk of the gas-phase carbon
is locked in CO molecules, and the CN and HCN densities depend 
on the rate at which carbon is
released from CO by cosmic-ray driven helium impact ionization.
Molecule formation is then ``ionization limited'' and occurs with an efficiency
proportional to the cosmic-ray ionization rate.
In dense clouds, an enhanced abundance of
atomic carbon is maintained at intermediate depths, where
charge transfer of C$^+$ with S is effective, and where
photo-destruction reduces the efficiency with which the carbon
atoms are removed in reactions with molecules.
At these intermediate depths the CN density  
is insensitive to $A_V$. However, 
the HCN densities increase with $A_V$ as the destructive FUV
photons are absorbed.

The C$^+$, C, CO, CN, and HCN
densities in the opaque cores depend on whether
the cores are in the ``low-ionization'' or 
``high-ionization'' phases that are possible 
for such gas. We present computations for 
gas in both phases.
The transition from the HIP to LIP occurs
at a critical value $n/\zeta_{-17}=10^3$ cm$^{-3}$,
consistent with previous findings.
The CN/HCN density ratio can become large $\gtrsim 1$
in the LIP, but remains small in the HIP.

For dense gas in Galactic molecular clouds, and for characteristic
cosmic-ray ionization rates, a large CN/HCN density ratio may 
be interpreted reliably as an indicator of molecule
formation in PDRs. In clouds exposed to enhanced fluxes
of cosmic-rays, as perhaps occurs in starburst galaxies,
a high CN/HCN ratio may alternatively be an indicator
of opaque clouds in the high-ionization phase.
Measurements of the C/CO ratio can be used to distinguish
between these two possibilities.

     
\section*{Acknowledgments}

We thank A. Dalgarno, A. Fuente, O. Gnat, E. Herbst, and C.F. McKee 
for discussions, and the referee for helpful comments 
and suggestions. This research is supported by the Israel 
Science Foundation, grant 221/03.



\clearpage

\begin{figure}
\plotone{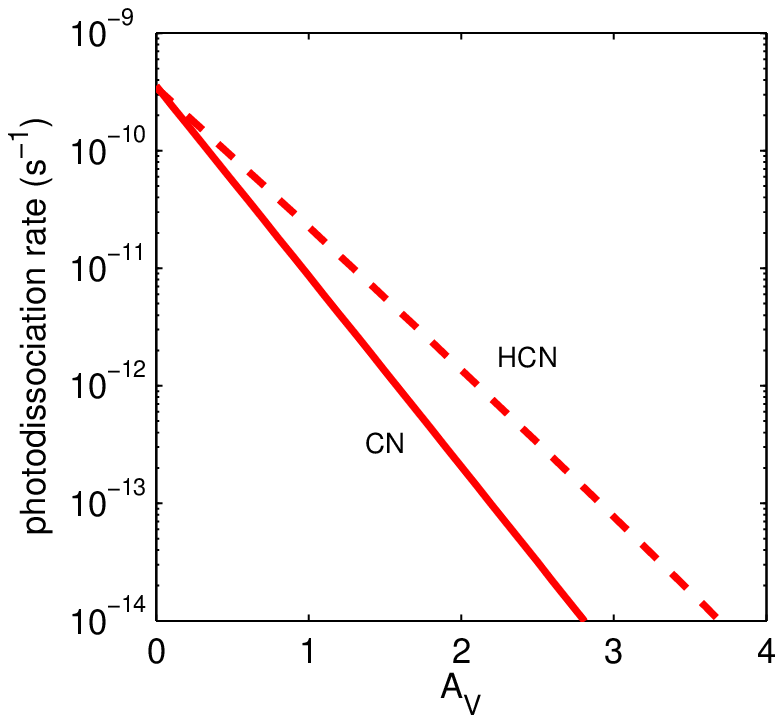}
\caption{Photodissociation rates of CN and HCN as functions of
visual extinction for a unit Draine field.} 
\label{prates}
\end{figure}

\clearpage

\begin{figure}
\includegraphics[]{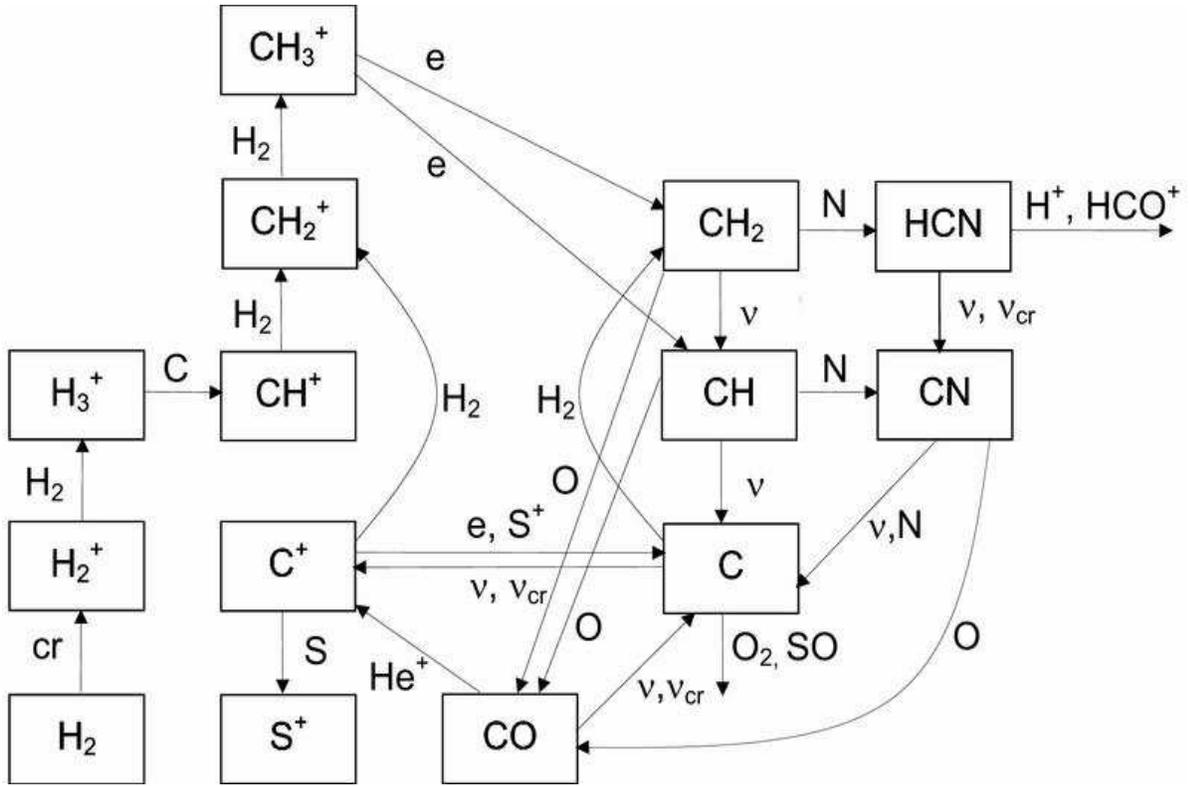}
\vspace{1in}
\caption{CN and HCN formation pathway \#1 via carbon hydride intermediates.} 
\label{path1}
\end{figure}

\clearpage
\begin{figure}
\includegraphics[]{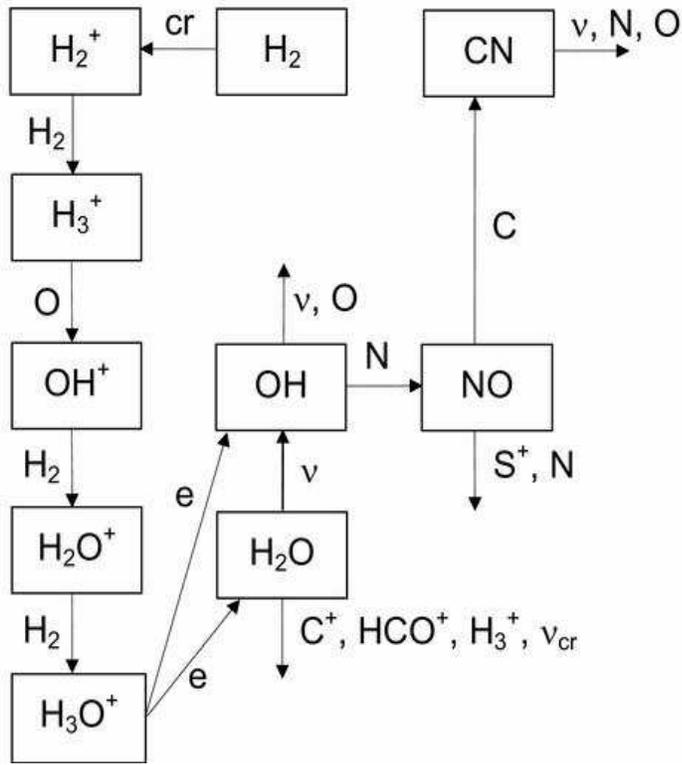}
\vspace{1in}
\caption{CN formation pathway \#2 via oxygen hydride intermediates.} 
\label{path2}
\end{figure}

\clearpage
\begin{figure}
\includegraphics[]{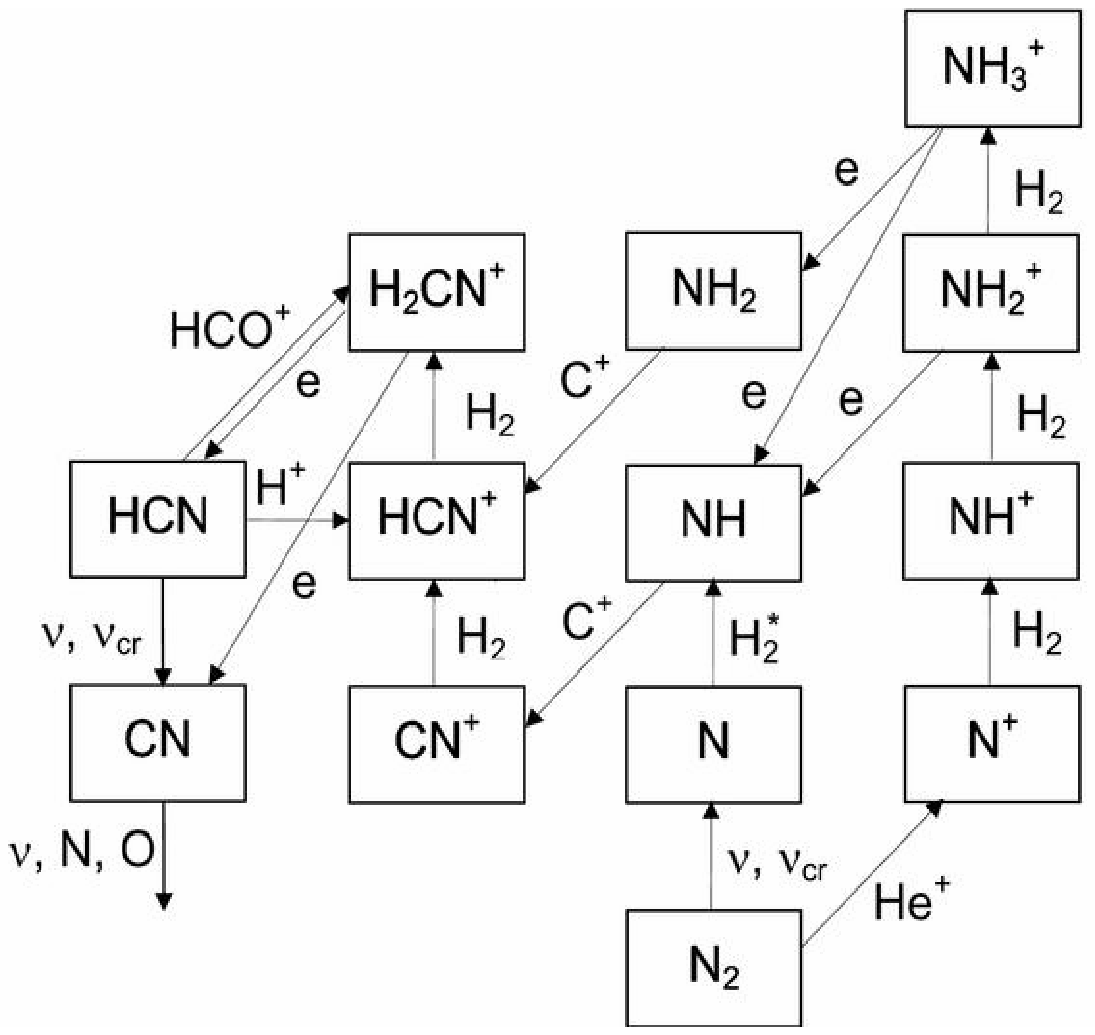}
\vspace{1in}
\caption{CN and HCN formation pathway \#3 via 
 nitrogen hydride intermediates.} 
\label{path3}
\end{figure}

\begin{figure}
\plotone{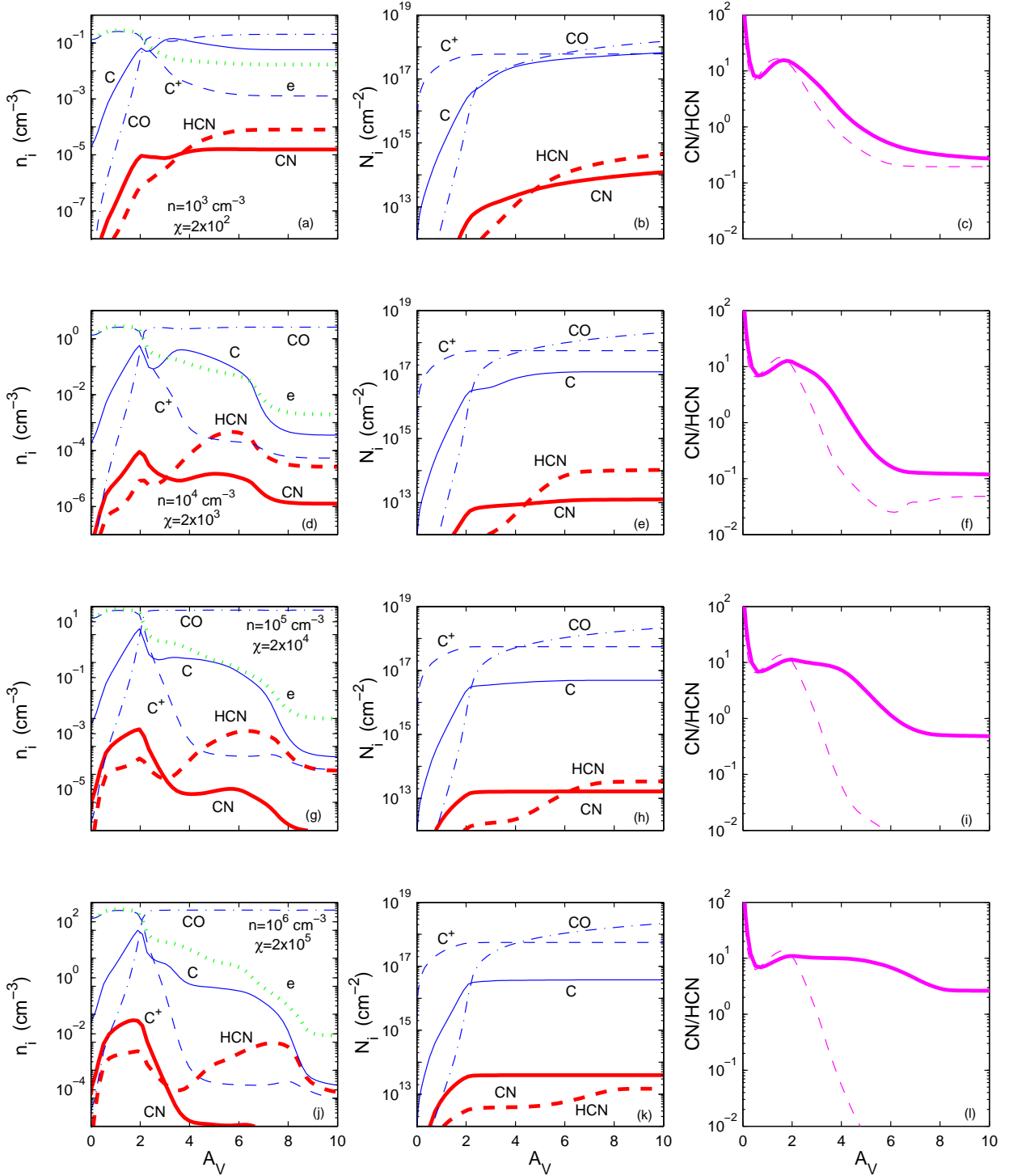}
\caption{Model results for $n$ equal to
$10^3$, $10^4$, $10^5$, and $10^6$ cm$^{-3}$ 
(from top to bottom) for constant ionization
parameter $\chi/n$ = 0.2 cm$^3$. 
The cosmic-ray ionization rate $\zeta=5\times 10^{-17}$~s$^{-1}$ in 
all four models.
Displayed profiles are for
C$^+$, C, CO, CN, HCN, and electrons, as functions
of~$A_V$. The lefthand panels display the volume densities.
The middle panels show the integrated column densities.
The righthand panels show the volume (dashed) and column (solid)
CN/HCN density ratios. The second row, with
$n=10^4$ cm$^{-3}$, $\chi=2\times 10^3$
is our ``reference model''.} 
\label{PS1}
\end{figure}

\begin{figure}
\plotone{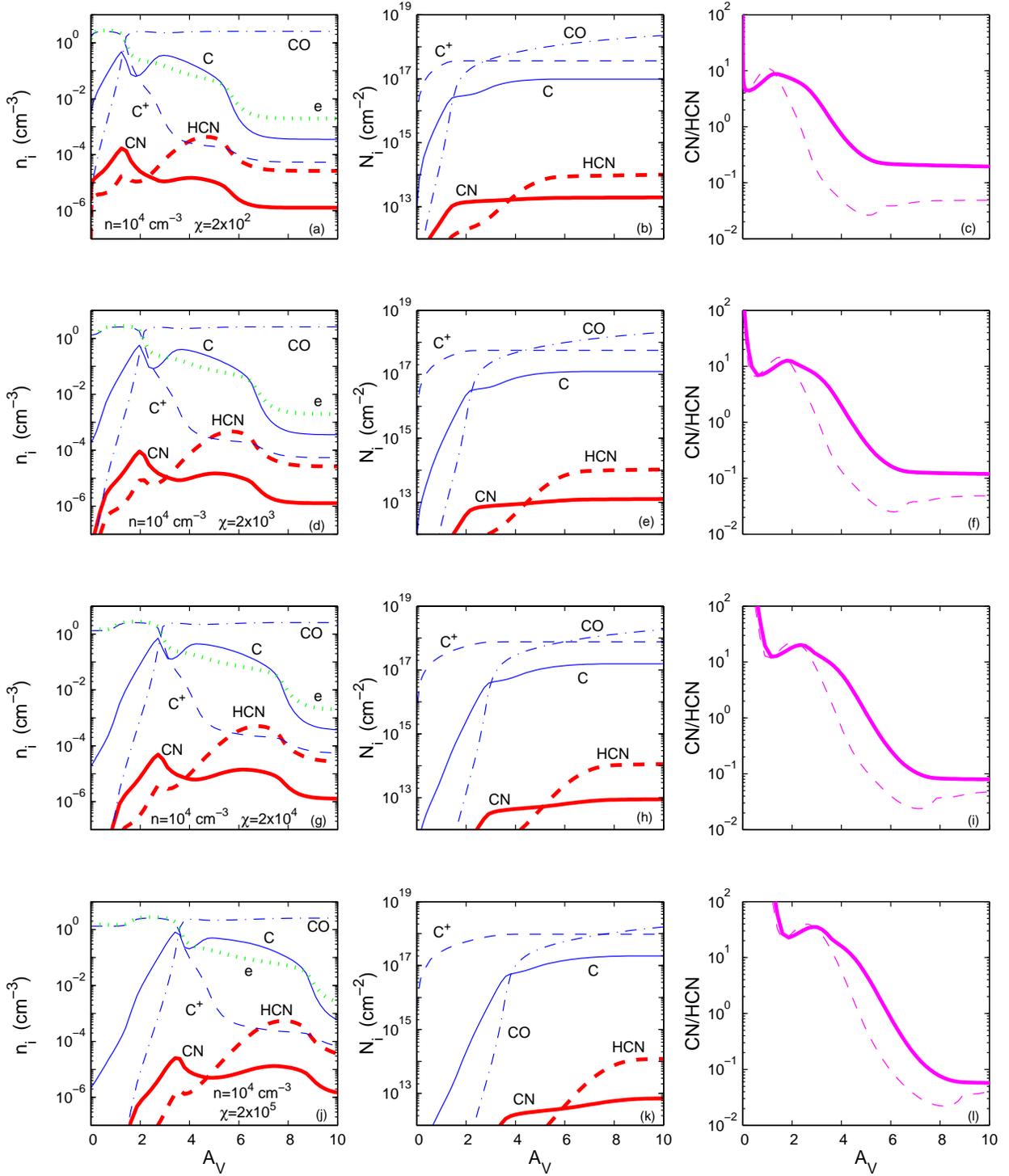}
\caption{Model results for $n=10^4$ cm$^{-3}$, and $\chi$ equal to
$2\times 10^2$, $2\times 10^3$, $2\times 10^4$, and $2\times 10^5$ cm$^{-3}$,
and $\zeta=5\times 10^{-17}$ s$^{-1}$. The second row is our reference model.}
\label{PS2}
\end{figure}

\begin{figure}
\plotone{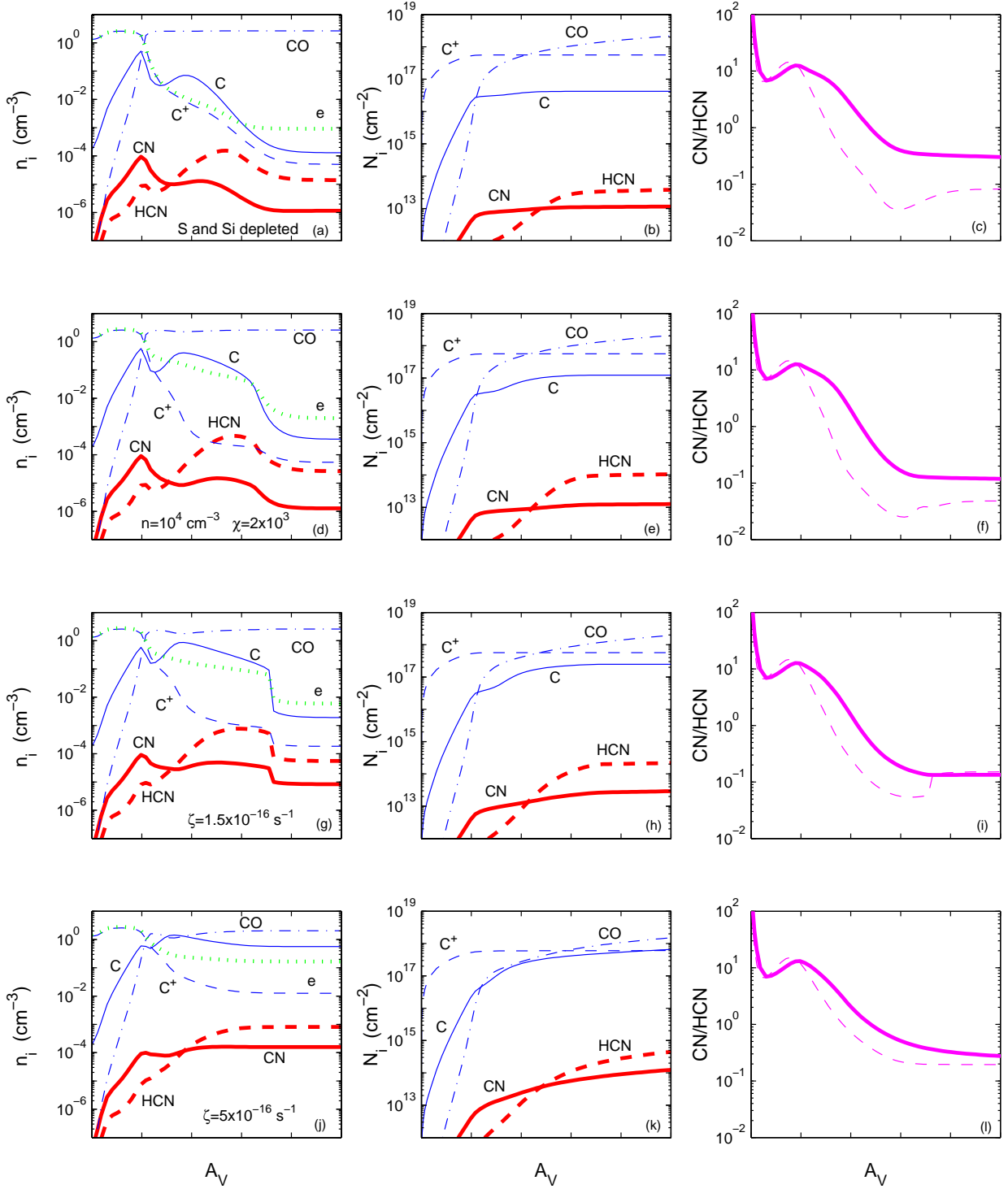}
\caption{Results for $n=10^4$ cm$^{-3}$, and $\chi=2\times 10^3$.
In the upper row the gas phase abundances of S and Si are reduced
by a factor of 100. The second row is our reference model,
with standard abundances and $\zeta=5\times 10^{-17}$ s$^{-1}$.
In the third and fourth rows, $\zeta$ is increased to 
$1.5\times 10^{-16}$ and $5\times 10^{-16}$ s$^{-1}$.}
\label{PS3}
\end{figure}

\begin{figure}
\includegraphics[width=8cm,height=22cm]{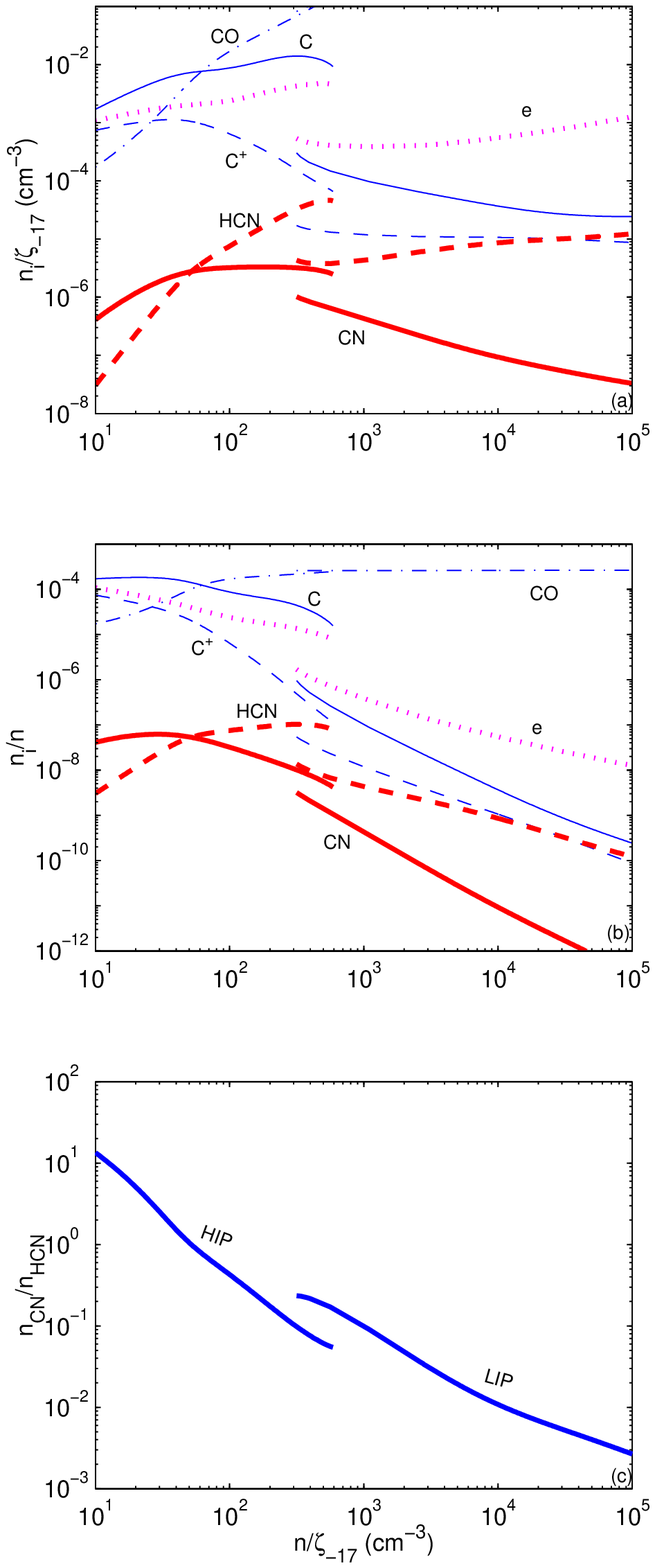}
\caption{Densities of CN, HCN, C$^+$, C, CO, and electrons,
in the cosmic-ray dominated cores, showing the high and low
ionization phases. Panels (a) and (b) show the densities 
$n_i/\zeta_{-17}$ and density fractions $n_i/n$, 
as functions of $n/\zeta_{-17}$, where
$\zeta_{-17}$ is the cosmic-ray ionization rate normalized to
$1.0\times 10^{-17}$ s$^{-1}$, and $n$ is the hydrogen (H$_2$)
gas density.  Panel~(c) displays the density ratio
$n_{\rm CN}/n_{\rm HCN}$ in the HIP and LIP.} 
\label{dark}
\end{figure}



\end{document}